\documentstyle[preprint,version2,aps]{revtex}

\begin{document}
\draft
\begin{title}
Nonuniversality in Imbibition
\end{title}
\author{P. B. Sunil Kumar\cite{Auth1} and
Debnarayan Jana\cite{Auth2}}
\begin{instit}
Raman Research Institute, Bangalore 560 080, India
\end{instit}
\begin{abstract}
We report an imbibition experiment in 2D random porous media in which
height - height correlation function grows with a nonuniversal exponent;
rather it depends on evaporation. We present an imbibition model which is
consistent with the experiment. The model also shows self-organisation
of the interface. To our knowledge this is the first model which explicitly
 allows for evaporation.
\end{abstract}
Recently there has been a lot of activity in understanding interface
growth phenomena~\cite{int}, particularly in deposition ~\cite{depo} and
imbibition ~\cite{exp,exp1}. Imbibition experiments in which a suspension
(for example coffee or ink) is imbibed into  paper have aroused a great
deal of interest in this area.  These simple experiments could help in
understanding the nature of diffusion through random porous media. This is
of importance in chromatography. They could also provide as good "table
top" systems to study pattern formation~\cite{pat,pat1}. In the
experiments a paper is fixed with the bottom end dipped into a suspension.
The suspension is imbibed into the pores of the paper by capillary action.
The fluid rises through the pores carrying the suspended particles with
it. An interface is formed by the wet front which rises steadily. There is
some amount of randomness present in the medium due to the blocking of the
pores. This randomness and evaporation of the fluid tries to pin this
wetting front. The motion of the wet front is impeded by the evaporation
rate, concentration of the suspension, the ratio of the size of the
suspended particles to the pore size and viscosity of the fluid. The front
stops moving when the fluid has completely evaporated. The roughness of this
interface is solely due to the disorder in the paper. If there is no blocking
of the pores the interface will be smooth at any rate of evaporation.
However the disorder experienced by the fluid is affected by the evaporation.
For smaller evaporation the fluid can go around the obstacles and thus not
{\it see} the blocks than when the evaporation is high.  As a result the
suspended particles carried by the fluid get deposited at this interface
causing a darkening of the boundary. It should be mentioned that this
imbibition case differs from the deposition models in two ways (i) there
is a time scale set by the evaporation (ii) the randomness in imbibition
is quenched and not fluctuating in time.

Till now most of the understanding of the interface growth phenomeona in
imbibition comes through computer simulations. So far no simulations have
been done taking care of evaporation, size of the particles and
concentration of the suspension explicitly. In the treatment by Amarlal
{\em et al} ~\cite{exp} evaporation was incorporated phenomenologicaly by
a steady increase $(\Delta p)$ at each time step in the probability $p$ of
blocking the pores. This increase drives the system into a percolation
threshold wherein they obtained a connected cluster of blocked pores.
This treatment assumes that the randomness is fluctuating in time.

The correlations of the fluctuations in the height difference between two
points $x$ and $x+l$ after the wetting front stops is given by

\begin{equation}
W(l)=\left<(h(x)-h(x+l))^2\right>_x^{1/2} \sim l^{\alpha}.
\end{equation}
The time correlation of the height growth at any point $x$ is defined as
\begin{equation}
C(\delta t)=\left<(h(t)-h(t+\delta t))^2\right>_t^{1/2} \sim \delta t^{\beta}
\end{equation}

In reference ~\cite{exp} the exponents $\alpha$ and $\beta$ obtained from
simulations are  $.63$ and $.68$ respectively. They claimed these
exponents to be universal.  In the experiment reported in reference ~\cite
{exp} also $\alpha$ was found to be $.65$.  But more experiments are
necessary to establish the universality of these exponents (i.e. the
independence of the exponents on various factors). And more theoretical
work as well as simulations taking account of these factors influencing
interface growth are also required.  Since in the imbibition experiments
described above the noise is due to the disorder present in the paper a
quenched randomness is more appropriate.

In this paper we present the results of a preliminary experimental study.
We find that the exponent $\alpha$ depends on the evaporation rate. A
model for the imbibition in 2D random porous media which takes care of the
effect of evaporation explicitly is described.  We have carried out a
simulation study of this model which gives results in good agreement with
our preliminary experiments.  We address ourselves to the static and
dynamic behaviour of the rough interface.

The experiments are carried out using Whatman No:1 filter paper as the
porous medium and ink as the suspension. The evaporation rate was varied
by changing the room humidity and temperature. The ink rises through the
paper and stops at a particular height. The darkening of the interface is
indicative of this stoppage of growth. The interface was then digitized
using a CCD camera and a frame grabber with a resolution of 260 pixels per
inch. Figure 1 shows the behavoir of $W(l)$ against $l$ for two different
values of evaporation rate.  The data are averaged over 10 experiments. We
find that the exponent $\alpha$ is very different for the two cases. This
shows that unlike the results of reference ~\cite{exp} $\alpha$ is not
universal.

To understand the dependence of the various parameters affecting the
growth of interface we present a model for studying the imbibition of a
liquid into 2D random porous media. Before discussing the model used in
this simulation let us try to understand the problem from a microscopic
point of view. In the experiments described above paper was used as the
random medium. At a microscopic level one can regard the paper as a
randomly disordered medium~\cite{exp} with a fixed probability $p$ for the
pores to be blocked. So, the interface growth phenomena is nothing but the
propagation of fluid particles through this disordered medium. The wetting
front of the fluid particles propagates due to the capillary forces. The
disorder in the medium and the evaporation tries to pin this growth.
Evaporation constantly decreases the number of fluid particles in the wetting
 front. This makes it more difficult for the fluid to pass the obstacles.
The front stops moving at a critical height due to this pinning. It is true
that the smaller the evaporation rate, larger will be the critical height.

In our model the porous medium is considered as a square lattice with
disorder being incorporated by blocking some cells randomly with a
constant probability $p$ ({\rm see figure 2}). The maximum capacity of
each  cell is fixed to $N_0$ number of fluid particles.  At time t=0, at
the bottom edge of the lattice  a horizontal line of wet cells with $N_0$
particles is created.  At t+1 the particles are imbibed into all unblocked
cells which are nearest neighbors to the wet region.  When a cell
transfers to its nearest neighbours the number of particles it contains
remains the same due to the source below. If a cell has more than one wet
nearest neighbour it gets particles from all of them subject to a maximum
number $N_0$.  At every time step, evaporation was explicitly modelled by
the loss of certain number of particles $n$ in the transfer. We also apply
the rule that every cell blocked or unblocked below a new wet cell become
wet as well \cite{exp} to avoid the presence of overhangs and islands.  We
use periodic boundary condition in $x$ by indentifying the cells at the
edge of the lattice.

This model is different from the model used previously\cite{exp}, in the
sense that here the effect of evaporation is incorporated in an explicit
way. Unlike the directed percolation models where the interface is pinned
by the connected cluster of blocks here it stops because it runs out of
fluid. The model is also different from the Eden growth in the sense that
all the sites in the boundary moves at the same time {\it i.e.} here the
growth process by itself does not cause interface roughness. The
concentration of the suspension and the pore size are incorporated in the
model in the following way. For a given
 concentration the number of fluid particles to the number of suspended
particles is fixed. Thus for a fixed pore size the change in concentration of
 the suspension or {\it Vice Versa} is same as changing the maximum number of
fluid particles $N_0$ which a cell can hold. Also it is possible to
include the effect of other factors like gravity, which can induce a bias
to the propagation of the fluid particles through the disordered medium.
The model can be used to study for example {\it thin layer chromatography}
wherein a mixture of different chemical compounds are made to diffuse
through a porous medium resulting in their separation.

The simulations are done on a lattice of length $6000$ lattice units. The
results are averaged over $500$ realisations. The simulations show
the existence of a crossover length $l_{x}$
such that the height - height correlation function $W \sim l^{\alpha}$ for
$l\ll~l_{x}$. Whereas for  $l\gg~l_{x}$ $W$ saturates to a constant value
$W_{sat}$ as shown in figure 3. This saturation width depends upon the
number $ n $ of the fluid particles evaporated as

\begin{equation}
W_{sat}~\sim (n)^{-\gamma}
\end{equation}

So, one can define a scaling form given by ~\cite{exp}

\begin{equation}
W(l,n)~\sim~ l^\alpha f(l n^{(\frac{\gamma}{\alpha})}) \label{scale}
\end{equation}

\noindent
where $f(x) \rightarrow$ a constant as $ x \rightarrow 0$ and $f(x)
\rightarrow x^{-\alpha}$ as $x\rightarrow \infty$.

We find that there are two critical values of evaporation $n_1$ and $n_2$
for a given value of $N_0$. These critical values change in such a way
that the ratios $n_1/N_0$ and $n_2/N_0$ remain constant.  For low values
of evaporation the mean height increases without stopping (regime I).
Above the critical evaporation $n_1$ the mean height stops after a finite
time.  For $n>n_1$ (regime II) the $log(W)\,\,-\,\,log(l)$ plots fall on
to a single curve through the scaling form  given in equation~\ref{scale}
(see figure 4).  We find in this region $\alpha=0\cdot 5$ and
$\gamma=3.0$.  Note that this is very different from the exponent obtained
in ~\cite{exp}. At the second critical evaporation rate $n_2$ this scaling
breaks down. For $n>n_2$ (regime III) the exponent $\alpha$ decreases
continuously with $n$ as can be seen from figure 3. The critical values
$n_1/N_0\,\,\,{\rm and}\,\,\,n_2/N_0$ decreases with $p$. For $p=.45$ we
get $n_1/N_0=.124$ and $n_2/N_0=.134$.

We can get more insight  by looking at how the height - height correlation
 exponent $\alpha$ changes with time. This is depicted in figure 5. At short
times $\alpha$ shows a rapid increase.  This region is independent of
evaporation. However the long time behaviour shows $\alpha$ saturating to
a value $\alpha_{sat}$.  This establishes the fact that the roughness of
the interface is controlled by evporation and shows that the reason for
dependence of $\alpha$ on $n$ is not due to the lack of time for the
interface to saturate. We see from figure 5 that for $n<n_1$ $\alpha$ is
greater than $.5$.  However in this low evaporation regime the front does
not get pinned but moves at a constant velocity. This dependence of $\alpha$
on evaporation is consistent with the experimental results described
above. However more experiments are necessary to confirm the smooth
dependence of $\alpha$ on evaporation predicted by the model. To establish
the universal dependence of $\alpha$ on evaporation more experiments using
different kinds of papers and suspensions are being pursued.

The dependence of the dynamical exponent $\beta$ on evaporation is the
same as that of $\alpha$. In figure 6 we show the behaviour of $ C(\delta
t) $ for various values of $n$ for a fixed values of $p$.  We see that in
regime I all the curves have a slope greater than  $.5$.  This slope
changes continuously to $.5$ in regime II.  In regime III the slope
decreases with $n$.

In this model we can apply  an external bias which could be present due to
gravity or anisotropy in the medium. This bias was incorporated in the
model by introducing a difference in the number of particles transfered to
the vertical and horizontal neighbours of a given cell. The different
regimes described above were also  observed in this case for a fixed value
of $n$ and $p$ as the bias was varied.

The correlation functions $W(l)$ and $C(\delta t)$ can be directly mapped
to the root mean square displacement in a random walk, with $l$ taking the
place of time in $W(l)$.  The critical behaviour at $n_2$  described above
is similar to the dynamical transition in biased random walk in a random
medium ~\cite{rev}, wherein the exponent changes continuously with the
applied bias is above a critical value.  Since the behaviour found with the
external bias and that due to evaporation are the same one can assume that
the evaporation effectively introduces a bias. The detailed nature of this
bias is however not clear.

In the model the particles have more than one path to reach a particular
cell. The effective number of paths available to reach a cell decreases
since increased evaporation suppresses the longer paths. For $n<n_1$ the
number of fluid particles that a cell loses through evaporation is more than
compensated by the inflow because of the many paths available. Hence this
regime becomes super diffusive. The value of exponents being greater than
$.5$ support this argument. As the number of paths become less the
particles get stuck at obstacles for a longer time inducing a transition
into a normal diffusive regime with exponents $.5$ ~\cite{kpz}. On further
increase of evaporation the system becomes subdiffusive.  In the case of
biased random walk this region is known to have exponents which change
continously with bias ~\cite{rev}.

The effect of evaporation on the roughness exponent is similar to the
dependence of exponents on amount of nonconservation in
cellular-automaton. These systems are known to exhibit self-organisation
{}~\cite{soc}. We establish the self-organisation of the interface in
imbibition by looking at the dependence of $W_{sat}$ on the system size
$L$. This is shown in fig.7. We find $W_{sat} \sim L^{0.12}$. A detailed
analysis of the effect of parameters on the self-organisation will be
published elsewhere.

In conclusion we have shown the breakdown of universality in imbibition
from direct experiments. A model for imbibition which includes the effect
of evaporation in a very natural way is presented. The model shows that
the static height - height correlation exponent $\alpha$ depends on the
 evaporation which is consistent with our experiments. The interface is shown
to exhibit self-organised criticality.

\noindent

\acknowledgements

We thank Somnath Bharadwaj, Supurna Sinha, Madan Rao and Jayanth  Banavar
for discussions and Gautam I. Menon for introducing us to imbibition. Our
thanks are due to Jayadev Rajagopal and S. Chanthrasekaran for their help
with the image acquisition and digitization. We acknowledge the referee for
suggestions.

 \figure{The experimental values of
height - height correlation function W(l) plotted against the distance of
separation l after the interface stopped growing. Points marked with *
fall onto a curve with exponent $.45 \pm .004$. For higher evaporation the
points (marked o ) correspond to an exponent $.67 \pm .004$}
\figure{Example of the multiple connectivity of the model for a 6X6
lattice. The blocked cells are shown black. At time 0 all the cells
corresponding to i=1 and j=1,6 are filled with $N_0$ particles. At each
time step the particles are transfered to nearest neighbours. Note that
the cell (4,4) gets particles from both (4,3) and (4,5). Also when a cell
(${\rm i}_{\rm s}$,j) is wet all the other cells (i,j) with ${\rm i}<{\rm
i}_{\rm s}$ are wet as well.} \figure{The simulated height - height
correlation function W(l) plotted against the distance of separation l
after the columns stopped growing. The length is measured in units of
lattice parameter. The parameters are $p=.45$, $n=18.1$ to $22.3$ in equal
intervals of $.6$. The saturation value of $W(l)$ decreases monotonically
with $n$. The simulations are done with a lattice of length 6000 applying
periodic boundary condition.} \figure{The $W(l)$ plotted against $l$ in
the scaling form defined in the text for the parameters given in figure 4}
\figure{The growth of height - height correlation exponent $\alpha$ as a
function of time for $p=.2$. The three regimes are marked I, II and III.
The lines are for $n=43,44.2,45$ respectively. The behaviour of
$\alpha_{sat}$ with evaporation $n$ in this three regimes is shown in the
inset.} \figure{The plot of dynamic correlation function $C(\delta t)$ for
various values of $\delta t$. The dotted line corresponds to a slope of
$.5$. Other parameters are same as figure 4.} \figure{The simulated height
- height correlation function W(l) plotted against the distance of
separation l after the columns stopped growing for system sizes
L=6000,4000,2000,500. p=.45, $n/N_0 =.126 .$} \end{document}